\documentclass[prd,aps,showpacs,nofootinbib,superscriptaddress]{revtex4-1}
\usepackage{graphicx}% Include figure files
\usepackage{dcolumn}% Align table columns on decimal point
\usepackage{bm}% bold math
\usepackage{amssymb}
\usepackage{amsfonts}
\usepackage{amsmath}
\usepackage{latexsym}
\usepackage{dsfont}
\usepackage{multirow}

\usepackage{color}

\definecolor{nv}{rgb}{0.1,0.1,0.6}
\definecolor{pr}{rgb}{0.2,0.1,0.5}
\definecolor{mg}{rgb}{0.4,0.0,0.4}

\newcommand{\beq}{\begin{equation}}
\newcommand{\eeq}{\end{equation}}
\newcommand{\beqy}{\begin{eqnarray}}
\newcommand{\eeqy}{\end{eqnarray}}
\newcommand{\beqyn}{\begin{eqnarray*}}
\newcommand{\eeqyn}{\end{eqnarray*}}

\newcommand{\bs}{\begin{slide}}
\newcommand{\es}{\end{slide}}
\newcommand{\bc}{\begin{center}}
\newcommand{\ec}{\end{center}}
\newcommand{\bmin}{\begin{minipage}}
\newcommand{\emin}{\end{minipage}}

\newcommand{\ud}{\mathrm{d}}

\newcommand{\bi}{\begin{itemize}}
\newcommand{\ei}{\end{itemize}}

\newcommand{\pure}{\text{pure}}
\newcommand{\phys}{\text{phys}}

\begin{document}

\preprint{APS/123-QED}

\title{Gauge symmetry and background independence:\\
Should the proton spin decomposition be path independent?}

\author{C\'edric Lorc\'e}
\email{lorce@ipno.in2p3.fr;C.Lorce@ulg.ac.be}
\affiliation{IPNO, Universit\'e Paris-Sud, CNRS/IN2P3, 91406 Orsay, France}
\affiliation{IFPA,  AGO Department, Universit\'e de Li\` ege, Sart-Tilman, 4000 Li\`ege, Belgium}

\date{\today}% It is always \today, today,

\begin{abstract}
Exploring the similarities between the Chen \emph{et al.} approach, where physical and gauge degrees of freedom of the gauge potential are explicitly separated, and the background field method, we provide an alternative point of view to the proton spin decomposition issue. We show in particular that the gauge symmetry can be realized in two different ways, and discuss the relations between the concepts of path dependence, Stueckelberg dependence and background dependence. Finally, we argue that path/Stueckelberg/background-dependent decompositions of the proton spin are in principle measurable and therefore physically meaningful.
\end{abstract}


\maketitle

\section{Introduction}

Until 2008, there were essentially two decompositions of the proton spin into quark/gluon and spin/orbital angular momentum (OAM) contributions. On the one hand, the Ji decomposition \cite{Ji:1996ek} 
\begin{equation}\label{Jidec}
\vec J=\underbrace {\int\ud^3x\,\psi^\dag\tfrac{1}{2}\vec\Sigma\psi}_{\vec S^q_\text{Ji}}+\underbrace{\int\ud^3x\,\psi^\dag(\vec x\times i\vec D)\psi}_{\vec L^q_\text{Ji}}+\underbrace{\int\ud^3x\,\vec x\times(\vec E^a\times\vec B^a)}_{\vec J^g_\text{Ji}}
\end{equation}
with $\vec D=-\vec\nabla-ig\vec A^a t^a$, is gauge invariant but does not provide any split of the gluon angular momentum $\vec J^g_\text{Ji}$ into spin and OAM contributions. On the other hand, the Jaffe-Manohar decomposition \cite{Jaffe:1989jz} 
\begin{equation}\label{JMdec}
\vec J=\underbrace {\int\ud^3x\,\psi^\dag\tfrac{1}{2}\vec\Sigma\psi}_{\vec S^q_\text{JM}}+\underbrace{\int\ud^3x\,\psi^\dag(\vec x\times\tfrac{1}{i}\vec\nabla)\psi}_{\vec L^q_\text{JM}}+\underbrace{\int\ud^3x\,\vec E^a\times\vec A^a}_{\vec S^g_\text{JM}}+\underbrace{\int\ud^3x\,E^{ia}(\vec x\times\vec\nabla) A^{ia}}_{\vec L^g_\text{JM}}
\end{equation}
provides a complete split into quark/gluon and spin/OAM contributions, but is not gauge invariant. In practice, the Jaffe-Manohar decomposition is then considered in a fixed gauge, casting doubts on its measurability and therefore physical relevance.

In 2008, Chen \emph{et al.} \cite{Chen:2008ag,Chen:2009mr} proposed a complete decomposition of the proton spin consistent with gauge symmetry 
\begin{equation}\label{Chendec}
\vec J=\underbrace {\int\ud^3x\,\psi^\dag\tfrac{1}{2}\vec\Sigma\psi}_{\vec S^q_\text{Chen}}+\underbrace{\int\ud^3x\,\psi^\dag(\vec x\times i\vec D_\pure)\psi}_{\vec L^q_\text{Chen}}+\underbrace{\int\ud^3x\,\vec E^a\times\vec A^a_\phys}_{\vec S^g_\text{Chen}}+\underbrace{\int\ud^3x\,E^{ia}(\vec x\times\vec{\mathcal D}^{ab}_\pure) A^{ib}_\phys}_{\vec L^g_\text{Chen}}
\end{equation}
with $\vec D_\pure=-\vec\nabla-ig\vec A^a_\pure t^a$ and $\vec{\mathcal D}^{ab}_\pure=-\delta^{ab}\vec\nabla-gf^{abc}\vec A^c_\pure$. The crucial assumption was the possibility to separate the gauge degrees of freedom from the physical degrees of freedom in the gauge potential $\vec A(x)=\vec A_\pure(x)+\vec A_\phys(x)$. In particular, they were able to define a gauge-invariant gluon spin contribution $\vec S^g_\text{Chen}$, in apparent contradiction with textbook claims \cite{Jauch,Berestetskii}. This approach reopened old controversies, triggered numerous studies and finally led to the conclusion that there exist in principle infinitely many possible decompositions of the proton spin. For a review of the recent developments, see Refs. \cite{Lorce:2012rr,Leader:2013jra}.

The Chen \emph{et al.} decomposition turned out to be a \emph{gauge-invariant extension} (GIE) \cite{Hoodbhoy:1998bt,Ji:2012gc} of the Jaffe-Manohar decomposition, in the sense that it is formally gauge invariant but gives the same results as the Jaffe-Manohar decomposition considered in a fixed gauge. Ji criticized the Chen \emph{et al.} approach, arguing that their notion of gauge invariance does not coincide with the ``usual textbook type'' \cite{Ji:2009fu,Ji:2010zza}. Moreover, Ji, Xu and Zhao  \cite{Ji:2012gc} wrote that ``the GIE of an intrinsically gauge-noninvariant quantity is not naturally gauge invariant'' and that ``GIE operators are in general unmeasurable. So far, the only example is offered in high-energy scattering in which certain partonic GIE operators may be measured.'' This kind of statement is pretty confusing and seems at first sight self-contradictory. Indeed, how can a measurable quantity be ``not naturally gauge invariant'' or, using Wakamatsu's words \cite{Wakamatsu:2013voa}, ``not a gauge-invariant quantity in a \emph{true} or \emph{traditional} sense''?

The aim of this paper is to clarify the meaning of gauge invariance and the reason for the existence of infinitely many (in principle measurable) decompositions. While the similarity between the Chen \emph{et al.} approach and the background field method has been noted in Refs. \cite{Zhang:2011rn,Lorce:2013gxa}, no further investigations have been made so far in the context of the proton spin decomposition. We therefore explore the relation between these two approaches and propose a new point of view. In particular, we show that the gauge symmetry can be realized in two different ways within the Chen \emph{et al.} approach. When discussing the gauge invariance of a quantity, it is therefore necessary to specify what type of gauge invariance is invoked to avoid confusion. This will clarify the meaning of the claims mentioned above and why the Jaffe-Manohar decomposition is measurable \emph{via} its GIEs.
\\

The plan of this paper is as follows. In section \ref{secII}, we briefly present both the Chen \emph{et al.} approach and the background field method. In section \ref{secIII}, we discuss the non-local intepretation of the Stueckelberg symmetry and provide a new interpretation based on the background field picture. In particular, we show that one can define two kinds of gauge transformations in the Chen \emph{et al.} approach. Accordingly, we discuss in section \ref{secIV} the weak and strong forms of gauge invariance, and argue that the strong form is actually not necessary for ensuring in principle the measurability. Finally, the status of the different proton spin decompositions is commented in section \ref{secV}, and we conclude the paper with section \ref{secVI}.

\section{Chen \emph{et al.} approach \emph{vs.} background field method}\label{secII}

\subsection{The Chen \emph{et al.} approach in a nutshell}\label{secIIA}

As mentioned in the introduction, Chen \emph{et al.} proposed, in the context of the proton spin decomposition, an approach where the physical and unphysical (or gauge) degrees of freedom in the gauge potential are explicitly separated. This idea is actually not new, as observed in Ref. \cite{Lorce:2013gxa}, since it has already been adopted in 1962 by Schwinger \cite{Schwinger:1962zz,Schwinger:1962fg}, followed by Arnowitt and Fickler \cite{Arnowitt:1962cv}, though in a different context. The same idea reappeared and was generalized several times \emph{e.g.} in Refs.~\cite{Goto:1966,Treat:1973yc,Treat:1975dz,Duan:1979,Duan:1984cb,Duan:1998um,Duan:2002vh,Fulp:1983bt,Kashiwa:1996rs,Kashiwa:1996hp}.   Nevertheless, we follow the terminology in the recent literature, and refer to it as the Chen \emph{et al.} approach. 

In the covariant form of the Chen \emph{et al.} approach \cite{Wakamatsu:2010cb}, the gauge potential is split into \emph{pure-gauge} and \emph{physical} contributions
\begin{equation}\label{decomposition}
A_\mu(x)=A^\pure_\mu(x)+A^\phys_\mu(x).
\end{equation}
By definition, the pure-gauge term does not contribute to the field strength
\begin{equation}\label{cond1}
F^\pure_{\mu\nu}=\partial_\mu A^\pure_\nu-\partial_\nu A^\pure_\mu-ig[A^\pure_\mu,A^\pure_\nu]=0
\end{equation}
and behaves like the gauge potential under gauge transformations
\begin{equation}\label{cond2}
A^\pure_\mu(x)\mapsto \tilde A^\pure_\mu(x)=U(x)[A^\pure_\mu(x)+\tfrac{i}{g}\partial_\mu]U^{-1}(x).
\end{equation}
It follows from these conditions that $A^\pure_\mu(x)$ is a pure-gauge term
\begin{equation}
A^\pure_\mu(x)=\tfrac{i}{g}\,U_\pure(x)\partial_\mu U^{-1}_\pure(x),
\end{equation}
where $U_\pure(x)$ is some unitary matrix behaving as
\begin{equation}
U_\pure(x)\mapsto\tilde U_\pure(x)=U(x)U_\pure(x)
\end{equation}
under gauge transformations. The physical field is responsible for the field strength
\begin{equation}
F_{\mu\nu}=\mathcal D^\pure_\mu A^\phys_\nu-\mathcal D^\pure_\nu A^\phys_\mu-ig[A^\phys_\mu,A^\phys_\nu],
\end{equation}
where $\mathcal D^\pure_\mu=\partial_\mu-ig[A^\pure_\mu,\quad]$ is the pure-gauge covariant derivative in the adjoint representation. The covariant derivative acting on the quark field reads $D^\pure_\mu=\partial_\mu-igA^\pure_\mu$. It is also easy to see that the physical field behaves like the field-strength tensor under gauge transformations
\begin{equation}
A^\phys_\mu(x)\mapsto \tilde A^\phys_\mu(x)=U(x)A^\phys_\mu(x)U^{-1}(x).
\end{equation}
Note that one always has the possibility to work in the gauge $A^\pure_\mu(x)=0$, where $A_\mu(x)=A^\phys_\mu(x)$.

The defining conditions \eqref{cond1} and \eqref{cond2} are actually not sufficient to specify a unique split of the gauge potential into pure-gauge and physical fields. Indeed, starting from some split $A_\mu(x)=A^\pure_\mu(x)+A^\phys_\mu(x)$, one can easily write another perfectly acceptable split $A_\mu(x)=\bar A^\pure_\mu(x)+\bar A^\phys_\mu(x)$, where
\begin{equation}\label{ST}
\begin{split}
\bar A^\pure_\mu(x)&=A^\pure_\mu(x)+C_\mu(x),\\
\bar A^\phys_\mu(x)&=A^\phys_\mu(x)-C_\mu(x),
\end{split}
\end{equation}
with $C_\mu(x)=\tfrac{i}{g}\,U_\pure(x)U^{-1}_S(x)[\partial_\mu U_S(x)]U^{-1}_\pure(x)$. The new pure-gauge field $\bar A^\pure_\mu(x)$ satisfies the conditions \eqref{cond1} and \eqref{cond2} provided that the matrix $U_S(x)$ is gauge invariant. Because of the similarity with the Stueckelberg mechanism, the transformation $\phi(x)\mapsto\bar\phi(x)$ is referred to as the Stueckelberg transformation \cite{Lorce:2012rr,Stoilov:2010pv}. In particular, while gauge transformations act on the left of $U_\pure(x)$, Stueckelberg transformations act on the right
\begin{equation}\label{Stueck}
\bar U_\pure(x)=U_\pure(x)U^{-1}_S(x).
\end{equation}
The set of Stueckelberg transformations forms a group which is a copy of the gauge group. Note that the standard gauge-invariant Lagrangians involve only the full gauge field $A_\mu(x)$, and so are trivially Stueckelberg invariant.

\subsection{The background field method in a nutshell}\label{secIIB}

The background field method has been introduced in 1967 by DeWitt \cite{DeWitt:1967ub,DeWitt:1967uc,DeWitt:1980jv,'tHooft:1975vy,Grisaru:1975ei,Boulware:1980av,Abbott:1980hw}. It has been extensively used in gravity and supergravity \cite{'tHooft:1973us,'tHooft:1974bx,Deser:1977nt,Abbott:1981ff,Petrov:2007xva,Petrov:2012qn},  and in both continuum and lattice gauge theories \cite{Dashen:1980vm,Abbott:1982jh,Luscher:1995vs,Freire:2000bq,Binosi:2009qm,Binosi:2012st}. A nice introduction to the background field method can be found in Ref. \cite{Abbott:1981ke}.

In the background field method, one introduces a \emph{background} (or classical or external or auxiliary) field $B_\mu(x)$, which satisfies some classical equation of motion, in the classical Lagrangian. This background field is not dynamical and is fixed  \emph{a priori}. The choice of a particular background is essentially a matter of convenience, namely the one that simplifies best the calculations. Only the difference between the two gauge potentials
\begin{equation}
Q_\mu(x)\equiv A_\mu(x)-B_\mu(x)
\end{equation}
is considered to be physical and meant to be quantized\footnote{Sometimes in the literature, like \emph{e.g.} in Ref. \cite{Grisaru:1975ei}, one adopts the equivalent point of view that the gauge field $A_\mu(x)$ is decomposed into a background field $B_\mu(x)$ and a quantum field $Q_\mu(x)$.}.

Since there are two gauge potentials, one can naturally define two kinds of gauge transformations. One can, for example, transform the dynamical and background fields independently
\begin{align}
A_\mu(x)&\mapsto U_A(x)[A_\mu(x)+\tfrac{i}{g}\partial_\mu]U^{-1}_A(x),&B_\mu(x)&\mapsto B_\mu(x),\label{AGT}\\
A_\mu(x)&\mapsto A_\mu(x),&B_\mu(x)&\mapsto U_B(x)[B_\mu(x)+\tfrac{i}{g}\partial_\mu]U^{-1}_B(x).\label{BGT}
\end{align}
It is actually more common to consider the following two kinds of gauge transformations
\begin{align}
A_\mu(x)&\mapsto U_a(x)[A_\mu(x)+\tfrac{i}{g}\partial_\mu]U^{-1}_a(x),&B_\mu(x)&\mapsto B_\mu(x),\label{aGT}\\
A_\mu(x)&\mapsto U_p(x)[A_\mu(x)+\tfrac{i}{g}\partial_\mu]U^{-1}_p(x),&B_\mu(x)&\mapsto U_p(x)[B_\mu(x)+\tfrac{i}{g}\partial_\mu]U^{-1}_p(x).\label{pGT}
\end{align}
The covariant derivatives $D^B_\mu=\partial_\mu-ig B_\mu$ and $\mathcal D^B_\mu=\partial_\mu-ig[B_\mu,\quad]$ are defined with respect to the background field. 

So, the introduction of a background field in a gauge theory allows one to define two kinds of gauge transformations. Similarly, the introduction of a background field in General Relativity
%\footnote{Note that a multi-connection approach in General Relativity has recently been proposed in Ref. \cite{Khosravi:2013kha}.}
 and String Theory allows one two define two kinds of point transformations \cite{Smolin:2005mq,Rickles:2008,Rozali:2008ex,Barenz:2012av}. The transformations \eqref{aGT} and \eqref{pGT} act in the same way on the dynamical fields, but differently on the background field.  In Eq. \eqref{pGT}, the background field $B_\mu(x)$ transforms like the corresponding dynamical field $A_\mu(x)$. This kind of transformation is therefore interpreted as a \emph{passive} transformation (\emph{i.e.} a change of coordinates), since a mere change of coordinate axes affects in the same way both dynamical and background fields. It is also known as background\footnote{There is some confusion in the literature since sometimes background gauge transformations refer to Eq. \eqref{BGT} instead of Eq. \eqref{pGT}.} (or classical) gauge transformation, since the quantum field undergoes simply a rotation $Q_\mu(x)\mapsto U_p(x)Q_\mu(x)U^{-1}_p(x)$. In Eq. \eqref{aGT}, the background field is not affected. This kind of transformations is therefore interpreted as an \emph{active} transformation (\emph{i.e.} a diffeomorphism), since only the physical, dynamical objects are modified.  It is also known as quantum gauge transformation, since the quantum field transforms more like a gauge field $Q_\mu(x)\mapsto U_a(x)[A_\mu(x)+\tfrac{i}{g}\partial_\mu]U^{-1}_a(x)-B_\mu(x)$. In the quantization procedure of the theory, it is the quantum gauge invariance which is broken by the gauge-fixing terms added to the classical Lagrangian. The background gauge invariance is preserved and simplifies many calculations, working as a bookkeeping aid.

\section{Interpretations of the Stueckelberg symmetry}\label{secIII}

\subsection{Non-local interpretation}\label{secIIIA}
 
In practical realizations of the Chen \emph{et al.} approach, one breaks explicitly the Stueckelberg symmetry by imposing an additional condition on the physical field. In their original work \cite{Chen:2008ag,Chen:2009mr}, Chen \emph{et al.} imposed the Coulomb constraint $\vec\nabla\cdot\vec A_\phys(x)=0$ in order to make contact with the famous Helmholtz decomposition of QED. Hatta \cite{Hatta:2011zs,Hatta:2011ku} imposed instead the light-front constraint $A^+_\phys(x)=0$ in order to make contact with the parton model of QCD. In principle, any Stueckelberg-fixing constraint is as good as any other one. Only for reasons of convenience would a particular constraint be preferred. 

Using the Stueckelberg-fixing constraint and the condition \eqref{cond1}, the pure-gauge and physical fields can be written explicitly as non-local expressions in terms of $A_\mu(x)$. As an example, one can write in many cases the matrix $U_\pure(x)$ as a Wilson line $\mathcal W_\mathcal C(x,x_0)=\mathcal P[e^{ig\int^x_{x_0}\ud s^\mu A_\mu(s)}]$, where $x_0$ is a fixed reference point and $\mathcal C$ is a path parametrized by $s^\mu(t)$. The pure-gauge and physical fields then read \cite{Hatta:2011ku,Lorce:2012ce}
\begin{align}
A^\pure_\mu(x)&=\tfrac{i}{g}\,\mathcal W_\mathcal C(x,x_0)\tfrac{\partial}{\partial x^\mu}\mathcal W_\mathcal C(x_0,x),\\
A^\phys_\mu(x)&=-\int^x_{x_0}\mathcal W_\mathcal C(x,s)F_{\alpha\beta}(s)\mathcal W_\mathcal C(s,x)\,\tfrac{\partial s^\alpha}{\partial x^\mu}\,\ud s^\beta.
\end{align}
These expressions are clearly non-local and path dependent. Changing the path does not affect the gauge potential $A_\mu(x)$, but determines its split into pure-gauge and physical contributions. The freedom in the choice of the path is therefore at the origin of the Stueckelberg symmetry. Namely, Stueckelberg-invariant quantities are path independent, whereas path-dependent quantities are Stueckelberg non-invariant. 

The Stueckelberg dependence is actually more general than the mere path dependence. Indeed, some Stueckelberg-fixing constraints, like \emph{e.g.} the Coulomb constraint $\vec\nabla\cdot\vec A_\phys(x)=0$ or the Lorenz constraint $\partial_\mu A^\mu_\phys(x)=0$, cannot be written simply in terms of Wilson lines, and are in this sense path independent \cite{Belinfante:1962zz,Mandelstam:1962mi,Rohrlich:1965,Yang:1985,Kashiwa:1994jn,Kashiwa:1997we}.  The corresponding explicit expressions for $A^\pure_\mu(x)$ and $A^\phys_\mu(x)$ in terms of $A_\mu(x)$ are nevertheless again non-local \cite{Zhang:2011rn}. For example, in the original Chen \emph{et al.} split in QED based on the Coulomb constraint, the pure-gauge and physical fields read explicitly
\beq\label{explicitChen}
\begin{split}
A^\pure_\mu(x)&=-\partial_\mu\frac{1}{\vec\nabla^2}\vec\nabla\cdot\vec A(x),\\
A^\phys_\mu(x)&=-\frac{1}{\vec\nabla^2}\partial^iF_{i\mu}(x),
\end{split}
\eeq
where the non-locality arises from the fact that the inverse Laplace operator is an integral operator
\begin{equation} \label{inverseLaplace}
\frac{1}{\vec\nabla ^2}f(\vec x)  \equiv -\frac{1}{4\pi}\int \ud^3x' \,\frac{f(\vec x')}{| \vec x - \vec x'|}.
\end{equation}
Stueckelberg non-invariant quantities are therefore intrinsically non-local, \emph{i.e.} cannot be written as local expressions in terms of $A_\mu(x)$. We do not feel that the non-locality is a serious issue here. Indeed, as already mentioned, one can always work in the gauge where $A^\pure_\mu(x)=0$ and therefore $A^\phys_\mu(x)=A_\mu(x)$, which means that the non-locality can be gauged away.

\subsection{Background interpretation}\label{secIIIB}

We discuss here a new, alternative interpretation of the Stueckelberg symmetry. As one can see from section \ref{secII}, the Chen \emph{et al.} approach and the background field method are pretty similar. Clearly, the pure-gauge field plays essentially the role of a background field, whereas the physical field is the natural object to be quantized. Stueckelberg dependence is therefore a particular type\footnote{Stueckelberg dependence and background dependence cannot be identified, because in the Chen \emph{et al.} approach one considers only the particular case of a flat background $F^\pure_{\mu\nu}=0$.} of background dependence. It also follows that one can in principle define two kinds of gauge transformations in the Chen \emph{et al.} approach. Comparing Eq. \eqref{cond2} with Eq. \eqref{pGT}, it is clear that the gauge transformations considered by Chen \emph{et al.} correspond to the passive or background gauge transformations
\begin{equation}
\begin{split}
A_\mu(x)&\mapsto U_p(x)[A_\mu(x)+\tfrac{i}{g}\partial_\mu]U^{-1}_p(x),\\
A^\pure_\mu(x)&\mapsto  U_p(x)[A^\pure_\mu(x)+\tfrac{i}{g}\partial_\mu]U^{-1}_p(x),\\
A^\phys_\mu(x)&\mapsto U_p(x)A^\phys_\mu(x)U^{-1}_p(x).
\end{split}
\end{equation}
Similarly to Eq. \eqref{aGT}, we define the active or quantum gauge transformations in the Chen \emph{et al.} approach as follows
\begin{equation}
\begin{split}
A_\mu(x)&\mapsto U_a(x)[A_\mu(x)+\tfrac{i}{g}\partial_\mu]U^{-1}_a(x),\\
A^\pure_\mu(x)&\mapsto A^\pure_\mu(x),\\
A^\phys_\mu(x)&\mapsto U_a(x)[A_\mu(x)+\tfrac{i}{g}\partial_\mu]U^{-1}_a(x)-A^\pure_\mu(x).
\end{split}
\end{equation}

It is easy to see that an active gauge transformation is a composition of a passive gauge transformation given by $U_p(x)=U_a(x)$ with a Stueckelberg transformation given by $U_S(x)=U^{-1}_\pure(x)U_a(x)U_\pure(x)$. Equivalently, combining an active gauge transformation given by $U_a(x)=U_\pure(x)U_S(x)U^{-1}_\pure(x)$ with the corresponding inverse passive gauge transformation given by $U_p(x)=U^{-1}_a(x)$, we simply recover the Stueckelberg transformation of Eq. \eqref{ST}. So, we can interpret Stueckelberg-dependent quantities as quantities sensitive to the difference between active and passive gauge transformations, whereas Stueckelberg-invariant quantities are blind to this difference. This new point of view clarifies the connection between gauge symmetry and Stueckelberg symmetry. In particular, we now easily understand why the Stueckelberg symmetry group is a simple copy of the gauge symmetry group.

Distinguishing gauge transformations from Stueckelberg transformations stresses the difference between two aspects of gauge symmetry. On the one hand, gauge symmetry allows one to set a component of the gauge field to zero. This is basically achieved by a (passive) gauge transformation. On the other hand, gauge symmetry allows one to decide which components of the gauge field are dynamical. This is basically achieved by a Stueckelberg transformation. Using a single gauge field, one is forced to do the two things at the same time. By fixing some component of the gauge field to zero, one determines automatically which components are dynamical. But if one deals with two gauge fields, it is possible to set one component of $A_\mu(x)$ to zero without determining which components are dynamical, \emph{i.e.} without deciding what is $A^\phys_\mu(x)$.

\section{Gauge invariance and measurability}\label{secIV}

\subsection{Weak and strong forms of gauge invariance}\label{secIVA}

When discussing gauge transformations, one usually does not specify whether these are considered as active transformations (\emph{i.e.} the system is modified) or passive transformations (\emph{i.e.} the reference axes are modified) \cite{Guay:2004zz}. The reason for this is that there is usually no way to distinguish active gauge transformations from passive gauge transformations. Similarly, in General Relativity there is \emph{a priori} no way to distinguish diffeomorphisms (active point transformations) from coordinate transformations (passive point transformations). As a matter of fact, physicists tend to think of gauge transformations as passive transformations, and therefore consider that gauge symmetry is just a mere redundancy of the mathematical description. On the contrary, mathematicians think of gauge transformations as active transformations, and so consider that gauge symmetry is a physical property of the system.

This is true as long as one deals only with dynamical fields. But when one introduces a background field into the game, like \emph{e.g.} in the Chen \emph{et al.} approach with $A^\pure_\mu(x)$, this equivalence does not hold anymore, and one should distinguish two forms of gauge invariance:
\begin{itemize}
\item \emph{Weak gauge invariance} refers simply to the invariance under passive gauge transformations;
\item \emph{Strong gauge invariance} refers to the invariance under both active and passive gauge transformations.
\end{itemize}
From the non-local point of view of section \ref{secIIIA}, weak gauge invariance is just the standard gauge invariance, while strong gauge invariance consists in the standard gauge invariance supplemented by the requirement that the expressions must be local (or equivalently Stueckelberg invariant). Path-dependent quantities, being non-local, can then only be gauge invariant in a weak sense.

Clearly, when Ji invokes the usual textbook gauge invariance, he refers to the strong form of gauge symmetry, since textbooks essentially deal with local quantities. The claim of Ji, Xu and Zhao  \cite{Ji:2012gc} that ``the GIE of an intrinsically gauge-noninvariant quantity is not naturally gauge invariant'' should therefore be understood as ``a non-local GIE is gauge invariant in a weak sense only''. Wakamatsu is less restrictive. He indeed writes in Ref. \cite{Wakamatsu:2013voa} that ``if a quantity in question is seemingly gauge-invariant but path-dependent, it is not a gauge-invariant quantity in a \emph{true} or \emph{traditional} sense, which in turn indicates that it may not correspond to genuine observable'', which is motivated by the claim in Refs. \cite{Belinfante:1962zz,Mandelstam:1962mi,Rohrlich:1965,Yang:1985} that path dependence is a reflection of gauge dependence. Wakamatsu therefore considers that physical quantities may be non-local, provided that they are path independent. This amounts to requiring something in between the weak and strong forms of gauge symmetry.

\subsection{Measurability}\label{secIVB}

It is always said that measurable quantities, \emph{i.e.} quantities that can be extracted from experimental data, must be gauge invariant. Since there exist two forms of gauge symmetry in the presence of a background field, one should clarify this statement.

It is clear that the weak form of gauge symmetry is in principle necessary for ensuring measurability, whereas the strong form of gauge symmetry is in principle sufficient. The latter is however not necessary. Indeed, high-energy scattering provides us with an example where certain non-local GIE operators can in practice be measured. The QCD factorization theorems \cite{Collins} allow one to approximate some cross sections by the convolution of a (process-independent) soft part, generally parametrized in terms of non-local parton distribution functions, with a perturbatively calculable (process-dependent) hard partonic cross section. The gauge invariance of the non-local parton distribution functions is ensured by the introduction of Wilson lines\footnote{More generally, one can introduce phase factors, which in some cases take the form of Wilson lines.}. 

The complete theory of an isolated system should in principle be background independent\footnote{Note however that at a quantum level, one is forced to specify a particular background. Ensuring background independence represents one of the biggest challenges in Quantum Gravity and String Theory \cite{Smolin:2005mq,Rickles:2008,Rozali:2008ex,Barenz:2012av,Gryb:2010ge,Hossain:2010wy,Fredenhagen:2011hm,Belot:2011fa}.},
 and so are QED and QCD. Then, why should one deal with path/Stueckelberg/background-dependent quantities? The answer is that these quantities appear in controlled expansion frameworks, like the twist expansion provided by the QCD factorization theorems. Background dependence enters the description when one makes an approximation by truncating the expansion.

One can interpret the background field as representing somehow a reference configuration. The choice of a particular background is essentially a matter of convenience. In General Relativity, any experiment is necessarily made in a certain frame, which can then be used as the natural background for the description of a system. In gauge theories, the situation is more complicated since the experiments do not determine a particular ``frame'' in internal or gauge space.  In theory, one can choose to work with any background, but in practice one should be able to relate the corresponding path/Stueckelberg/background-dependent quantities to experimental results, \emph{i.e.} one needs to rely on a controlled expansion framework, and very few are actually available. This is the reason why Ji, Xu and Zhao \cite{Ji:2012gc} claimed that the ``GIE operators are in general unmeasurable''.

Coming back to the example of high-energy scattering, the shape of the Wilson lines is \emph{a priori} arbitrary, but since one is working with a truncated twist expansion, the Wilson lines are actually forced to run essentially along the light-front direction. In other words, parton distribution functions with Wilson lines running along the light-front directions contribute at a given order in the twist expansion, whereas parton distribution functions with other kinds of Wilson lines contribute at several (if not all) orders in the twist expansion. So, even if the actual measured cross sections do not depend on a particular choice of path, the extraction of parton distributions at a given twist determines uniquely the shape the Wilson lines. Working at a given twist then amounts to working with a background field such that $A^+_\pure(x)=A^+(x)$, \emph{i.e.} with the natural Stueckelberg-fixing constraint $A^+_\phys(x)=0$.

\subsection{Classification of the measurable quantities}\label{secIVC}

Measurable quantities can be sorted into two categories:
\begin{itemize}
\item \emph{Observables} are gauge-invariant quantities in a strong sense. They are local (\emph{i.e.} path/Stueckelberg/background independent) and can in principle be measured without relying on any expansion framework;
\item \emph{Quasi-observables} are gauge-invariant quantities in a weak sense. They are non-local (\emph{i.e.} Stueckelberg/background dependent) and can only be measured within an expansion framework.
\end{itemize}
According to this terminology, cross-sections are clearly observables. On the contrary, parton distribution functions are quasi-observables, since they are non-local quantities involving path-dependent Wilson lines. Contrary to observables, quasi-observables generally depend on the regularization scheme and the factorization scale.

Interestingly, some \emph{moments} of parton distribution functions turn out to be observables, because they can be associated with a local gauge-invariant operator appearing in the standard operator product expansion. For example, the quark helicity distribution defined as
\begin{equation}
\Delta q(x)=\tfrac{1}{2}\int\frac{\ud z^-}{2\pi}\,e^{ixP^+z^-}\langle P,S|\overline\psi(-\tfrac{z^-}{2})\gamma^+\gamma_5\mathcal W(-\tfrac{z^-}{2},\tfrac{z^-}{2})\psi(\tfrac{z^-}{2})|P,S\rangle,
\end{equation}
with $\mathcal W(-\tfrac{z^-}{2},\tfrac{z^-}{2})$ a straight Wilson line running along the light-front direction between the points $-\tfrac{z^-}{2}$ and $\tfrac{z^-}{2}$, is obviously a quasi-observable. However, its first $x$-moment $\int^1_{-1}\ud x \,\Delta q(x)$ coincides with the quark axial charge $\Delta q$ defined as
\begin{equation}
2\Delta q \,S^\mu=\langle P,S|\overline\psi(0)\gamma^+\gamma_5\psi(0)|P,S\rangle,
\end{equation}
which is clearly an observable. 

Similarly, the gluon helicity distribution defined as
\begin{equation}\label{Deltag}
\Delta g(x)= \frac{i}{x P^+}\int \frac{\ud z^-}{2\pi}\,e^{ixP^+z^-}
\langle P, S| 2\text{Tr}[F^{+\alpha}(-\tfrac{z^-}{2}) \mathcal W(-\tfrac{z^-}{2},\tfrac{z^-}{2})\tilde{F}^{+}_{\phantom{+}\alpha}(\tfrac{z^-}{2})\mathcal W(\tfrac{z^-}{2},-\tfrac{z^-}{2})]| P, S \rangle,
\end{equation}
where the dual field-strength tensor is given by $\tilde{F}^{\mu\nu} = \tfrac{1}{2}\,\epsilon^{\mu\nu\alpha\beta}F_{\alpha\beta}$ with $\epsilon_{0123}=1$, is also a quasi-observable. Contrary to the quark case, its first $x$-moment $\Delta g\equiv\int^1_0\ud x \,\Delta g(x)$ remains a quasi-observable, since it cannot be be written as a local gauge-invariant expression (there is \emph{e.g.} no corresponding operator in the operator product expansion), in agreement with the textbook claim that there exists no local gauge-invariant expression for the gluon spin. The quantity $\Delta g$ is measurable but, being a quasi-observable, has a delicate physical interpretation. In the light-front gauge $A^+(x)=0$ with antisymmetric boundary condition\footnote{We use the Cauchy principal-value prescription for the $1/x$ factor.} $A_\mu(+\infty^-)=-A_\mu(-\infty^-)$, the expression for $\Delta g$ becomes local and coincides with the Jaffe-Manohar (infinite-momentum frame) expression for the gluon longitudinal spin \cite{Jaffe:1989jz}
\begin{equation}
\Delta g =\frac{1}{2P^+}\,\langle P,  S |2\text{Tr}[F^{1+}(0)A^2(0) -F^{2+}(0)A^1(0)]| P, S   \rangle \big|_{A^+=0\,\text{+B.C.}}.
\end{equation}
Accordingly, $\Delta g$ is interpreted as a measure of the gluon longitudinal spin in the light-front gauge. Using the Chen \emph{et al.} approach, Hatta \cite{Hatta:2011zs} showed that this interpretation can be made gauge invariant. It is sufficient to fix the Stueckelberg symmetry with the light-front constraint $A^+_\phys(x)=0$ and antisymmetric boundary condition $A^\phys_\mu(+\infty^-)=-A^\phys_\mu(-\infty^-)$, while preserving the (weak) gauge invariance. In this case, the physical field reads explicitly
\begin{equation}
A^\mu_\phys(x)=-\int\ud z^-\,\tfrac{1}{2}\,\epsilon(z^--x^-)\,\mathcal W(x^-,z^-)F^{+\mu}(z^-)\mathcal W(z^-,x^-),
\end{equation}
where $\epsilon(z^--x^-)$ is the sign function, and one can rewrite Eq. \eqref{Deltag} in the (seemingly) local, gauge-invariant form
\begin{equation}
\Delta g =\frac{1}{2P^+}\,\langle P,  S |2\text{Tr}[F^{1+}(0)A^2_\phys(0) -F^{2+}(0)A^1_\phys(0)]| P, S   \rangle.
\end{equation}
This is simply the expression for the gluon longitudinal spin in the (light-front) GIE extension of the Jaffe-Manohar decomposition, also known as the Hatta decomposition \cite{Hatta:2011ku}.

\section{Physically relevant proton spin decompositions}\label{secV}

We now come back to the proton spin decompositions mentioned in the introduction and discuss their physical relevance.

In the Ji decomposition \eqref{Jidec}, all the contributions correspond to observables, following the terminology introduced in section \ref{secIVC}. In other words, the Ji decomposition has the advantage of being path/Stueckelberg/background independent. Accordingly, as shown by Ji \cite{Ji:1996ek}, the various contributions can be expressed in terms of energy-momentum and axial form factors, which can in turn be expressed in terms of moments of generalized parton distributions.

The Jaffe-Manohar decomposition \eqref{JMdec} considered in a certain gauge and its corresponding GIE \eqref{Chendec} are physically equivalent, in the sense that they give the same physical result. They represent actually an infinite set of decompositions, since one has to specify also which gauge or Stueckelberg constraint is used. The original Chen \emph{et al.} decomposition \cite{Chen:2008ag,Chen:2009mr} in QCD is based on the (generalized) Coulomb constraint $\vec{\mathcal D}^{ab}_\pure\cdot\vec A^b_\phys=0$, and therefore gives the same physical results as the Jaffe-Manohar decomposition in the Coulomb gauge. Alternatively, the Hatta decomposition \cite{Hatta:2011ku} has the same generic structure as Eq. \eqref{Chendec} but is based on the light-front constraint $A^+_\phys=0$, and therefore gives the same physical results as the Jaffe-Manohar decomposition in the light-front gauge. All these decompositions are Stueckelberg/background dependent, but are in principle measurable. 

Wakamatsu \cite{Wakamatsu:2013voa} favors the original Chen \emph{et al.} decomposition (or equivalently the Jaffe-Manohar decomposition in the Coulomb gauge) because it is path independent, contrary to the Hatta decomposition (or equivalently the Jaffe-Manohar decomposition in the light-front gauge). But from a practical point of view, it is the latter decomposition which can be experimentally accessed thanks to the QCD factorization theorems and the twist expansion. In other words, within the infinite set of decompositions, it is the Hatta decomposition (or equivalently the Jaffe-Manohar decomposition in the light-front gauge) which is physically meaningful, in the sense that it is in practice measurable\footnote{Note that even if the parton distribution related to the Jaffe-Manohar orbital angular momentum has been identified \cite{Hatta:2011ku,Lorce:2011kd,Lorce:2011ni,Lorce:2012ce}, it remains a challenge to find the actual process allowing for its extraction.}.

\section{Conclusion}\label{secVI}

The decomposition of the proton spin proposed by Chen \emph{et al.} reopened many old controversies. It turned out that infinitely many similar decompositions can be considered, related to the fact that the Chen \emph{et al.} approach is intrinsically non-local. In particular, it has been argued that this lack of uniqueness is a signature that the formal gauge invariance of the Chen \emph{et al.} decomposition is not the genuine (textbook) gauge invariance, casting doubts on its measurability and therefore physical relevance.

In order to clarify the situation, we discussed in more detail the origin of the non-uniqueness embodied by the Stueckelberg transformations, and the relation with the gauge symmetry. Exploring the similarities between the Chen \emph{et al.} approach and the background field method, we proposed a new point of view and interpretation of the Stueckelberg symmetry. In particular, we showed that there actually exist two forms of gauge invariance in the Chen \emph{et al.} approach: a weak gauge invariance (\emph{i.e.} invariance under passive transformations only) and a strong gauge invariance (\emph{i.e.} invariance under both active and passive transformations). We argued that the latter form is too strong a requirement for ensuring in principle measurability, as illustrated by parton physics in high-energy experiments.

We then concluded that path/Stueckelberg/background-dependent decompositions of the proton spin are perfectly measurable, provided that there exists a suitable expansion framework. The QCD factorization theorems provide us with such an example and confirm the physical relevance of path/Stueckelberg/background-dependent decompositions of the proton spin.

\section*{Acknowledgements}

In this study, I greatly benefited from numerous discussions with E. Leader and  M. Wakamatsu. This work was supported by the P2I (``Physique des deux Infinis'') network and by the Belgian Fund F.R.S.-FNRS \emph{via} the contract of Charg\'e de recherches.

\end{document}